\documentclass[12pt]{article}
\usepackage{bm,amsmath,amssymb,graphicx}

\begin{document}
\pagenumbering{arabic}
\begin{titlepage}

\title{On the vDVZ discontinuity in massive conformal gravity}

\author{F. F. Faria$\,^{*}$ \\
Centro de Ci\^encias da Natureza, \\
Universidade Estadual do Piau\'i, \\ 
64002-150 Teresina, PI, Brazil}

\date{}
\maketitle

\begin{abstract}
By taking the massless limit of linearized massive conformal gravity coupled 
to a source, we show that the theory is free from the vDVZ discontinuity. 
This result is confirmed when we take the massless limit of the gravitational 
potential of the theory, which is shown to be finite at the origin.
\end{abstract}

\thispagestyle{empty}
\vfill
 \par
\bigskip
\noindent * felfrafar@hotmail.com \par
\end{titlepage}
\newpage


\section{Introduction}


Recently, it has been explicitly shown that the classical theory of massive 
conformal gravity (MCG) has a negative energy massive tensor field 
\cite{Faria1}, which is a common feature of theories of gravity with fourth 
order derivative terms in their actions \cite{Stelle1}. The problem with 
such field is that it can produce instabilities on the classical solutions of 
the theory \cite{Smilga}. In addition, the theory may present a discontinuity 
in the limit as the mass of the field goes to zero. 

At the quantum level, the corresponding quantum state of a negative energy 
field is taken to have negative energy or negative norm \cite{Stelle2}. The 
requirement that the theory be renormalizable, makes it necessary to choose 
the negative norm ghost state over the negative energy state. This choice 
spoil the unitarity of the $S$-matrix unless the position of the ghost pole 
is gauge-dependent \cite{Antoniadis}. 

In quantum MCG, the negative energy massive tensor field gives rise to a 
massive spin-$2$ ghost state. The presence of this ghost state ensures the 
renormalizability of the theory \cite{Faria1,Faria2}. In addition, it has 
recently been shown that the position of the pole of the MCG ghost state 
depends on the conformal gauge \cite{Faria3}, which means that the theory 
is not only renormalizable but also unitary. Hence, MCG is a consistent 
theory of quantum gravity.   

The aim of this work is to verify part of the consistency of classical 
MCG. Due to the extent and complexity of the calculations on the stability 
of the classical MCG solutions, we will leave them for future works. Here 
we focus on studying the continuity of the theory in the massless limit. 
In Sect. \ref{sec2} we describe MCG in the presence of matter fields. In Sect. 
\ref{sec3} we analyze the particle content of the theory. In Sect. \ref{sec4} 
we take the massless limit of the theory and check whether it is continuous or 
not. In Sect. \ref{sec5} we evaluate the relation between the continuity of 
the theory and the behavior of its gravitational potential at the origin. 
Finally, in Sect. \ref{sec6} we present our conclusions.  


\section{Massive conformal gravity}
\label{sec2}


Let us consider the total MCG action\footnote{Here we consider units in which 
$c=\hbar=1$.} \cite{Faria4}
\begin{equation}
S_{\textrm{tot}} = 
\frac{1}{k^2}\int{d^{4}x} \, \sqrt{-g}\bigg[\varphi^{2}R 
+ 6\partial^{\mu}\varphi\partial_{\mu}\varphi - \frac{1}{2m^2} 
C^{\alpha\beta\mu\nu}C_{\alpha\beta\mu\nu} \bigg] 
+ \int{d^{4}x}\mathcal{L}_{m}, 
\label{1}
\end{equation}
where $k^2$ is a constant whose value will be determined in Sec. \ref{sec5}, 
$m$ is a constant with dimension of mass, $\varphi$ is a scalar field called 
dilaton,
\begin{equation}
C^{\alpha\beta\mu\nu}C_{\alpha\beta\mu\nu} = R^{\alpha\beta\mu\nu}
R_{\alpha\beta\mu\nu} - 4R^{\mu\nu}R_{\mu\nu} + R^2 
+ 2\left(R^{\mu\nu}R_{\mu\nu} - \frac{1}{3}R^{2}\right)
\label{2}
\end{equation}
is the Weyl tensor squared, $R^{\alpha}\,\!\!_{\mu\beta\nu}$ is the 
Riemann tensor, $R_{\mu\nu} = R^{\alpha}\,\!\!_{\mu\alpha\nu}$ is the 
Ricci tensor, $R = g^{\mu\nu}R_{\mu\nu}$ is the scalar curvature, and 
$\mathcal{L}_{m} = \mathcal{L}_{m}(g_{\mu\nu},\Psi)$ is the Lagrangian 
density of the matter field $\Psi$. Besides being invariant under coordinate 
transformations, the action (\ref{1}) is also invariant under the conformal 
transformations
\begin{equation}
\tilde{g}_{\mu\nu}=e^{2\theta(x)}\,g_{\mu\nu}, \ \ \ \ \
\tilde{\varphi}=e^{-\theta(x)} \varphi, \ \ \ \ \
\tilde{\mathcal{L}}_{m} = \mathcal{L}_{m},
\label{3}
\end{equation}
where $\theta(x)$ is an arbitrary function of the spacetime coordinates. 

It is worth noticing that under the transformations 
\begin{equation}
\hat{g}_{\mu\nu} = \varphi^{2}g_{\mu\nu}, \ \ \ \ \ \hat{\varphi} = 1, 
\label{4}
\end{equation}
the action (\ref{1}) takes the form
\begin{equation}
\hat{S}_{\textrm{tot}} = 
\frac{1}{k^2}\int{d^{4}x} \, \sqrt{-\hat{g}}\bigg[ \hat{R} 
- \frac{1}{2m^2} \hat{C}^{\alpha\beta\mu\nu}\hat{C}_{\alpha\beta\mu\nu} \bigg] 
+ \int{d^{4}x}\hat{\mathcal{L}}_{m}. 
\label{5}
\end{equation}
In addition, the MCG line element 
$ds^2 = (\varphi^{2}g_{\mu\nu})dx^{\mu}dx^{\nu}$ reduces to the general 
relativistic line element  
\begin{equation}
d\hat{s}^2 = \hat{g}_{\mu\nu}dx^{\mu}dx^{\nu},
\label{6}
\end{equation}
and the MCG geodesic equation
\begin{equation}
\frac{d^{2}x^{\lambda}}{d\tau^2} + \Gamma^{\lambda}\,\!\!_{\mu\nu}
\frac{dx^{\mu}}{d\tau}\frac{dx^{\nu}}{d\tau} +\frac{1}{\varphi}
\frac{\partial\varphi}{\partial x^{\rho}} \left( g^{\lambda\rho} + 
\frac{dx^{\lambda}}{d\tau}\frac{dx^{\rho}}{d\tau}\right) = 0
\label{7}
\end{equation}
reduces to the general relativistic geodesic equation
\begin{equation}
\frac{d^{2}x^{\lambda}}{d\tau^2} + \hat{\Gamma}^{\lambda}\,\!\!_{\mu\nu}
\frac{dx^{\mu}}{d\tau}\frac{dx^{\nu}}{d\tau} = 0,
\label{8}
\end{equation}
where
\begin{equation}
\Gamma^{\lambda}\,\!\!_{\mu\nu} = \frac{1}{2}g^{\lambda\rho}\left( 
\partial_{\mu}g_{\nu\rho} + \partial_{\nu}g_{\mu\rho} 
- \partial_{\rho}g_{\mu\nu} \right)
\label{9}
\end{equation}
is the Levi-Civita connection. Thus, in the absence of matter, MCG is 
equivalent to the Einstein-Weyl gravity. In the presence of matter, however, 
the two theories are not equivalent, since the dilaton field $\varphi$ 
reappears in the transformed matter Lagrangian density 
$\hat{\mathcal{L}}_{m} = \mathcal{L}_{m}(\varphi^{-2}\hat{g}_{\mu\nu},\Psi)$.

Since the integral of the Euler density
\begin{equation}
E = R^{\alpha\beta\mu\nu}R_{\alpha\beta\mu\nu} - 4R^{\mu\nu}R_{\mu\nu} + R^2
\label{10}
\end{equation} 
is topologically invariant, we can write (\ref{1}) as
\begin{equation}
S_{\textrm{tot}} = \frac{1}{k^2}\int{d^{4}x} \, \sqrt{-g}\bigg[
\varphi^{2}R + 6\partial^{\mu}\varphi \partial_{\mu}\varphi 
- \frac{1}{m^2} \left( R^{\mu\nu}R_{\mu\nu} - \frac{1}{3}R^{2}\right) \bigg] 
+ \int{d^{4}x}\mathcal{L}_{m}.
\label{11}
\end{equation}
By varying (\ref{11}) with respect to $g^{\mu\nu}$ and $\varphi$, we obtain 
the MCG field equations
\begin{equation}
\varphi^{2}G_{\mu\nu} +  6 \partial_{\mu}\varphi\partial_{\nu}\varphi 
- 3g_{\mu\nu}\partial^{\rho}\varphi\partial_{\rho}\varphi 
+ g_{\mu\nu} \nabla^{\rho}\nabla_{\rho} \varphi^{2}
- \nabla_{\mu}\nabla_{\nu} \varphi^{2} - \frac{1}{m^2}W_{\mu\nu} 
= \frac{1}{2}k^2T_{\mu\nu},
\label{12}
\end{equation}
\begin{equation}
\left(\nabla^{\mu}\nabla_{\mu} - \frac{1}{6}R \right) \varphi = 0,
\label{13}
\end{equation}
where
\begin{equation}
W_{\mu\nu} = \nabla^{\alpha}\nabla^{\beta}C_{\mu\alpha\nu\beta} 
-\frac{1}{2} R^{\alpha\beta}C_{\mu\alpha\nu\beta} 
\label{14}
\end{equation}
is the Bach tensor,
\begin{equation}
G_{\mu\nu} = R_{\mu\nu} - \frac{1}{2}g_{\mu\nu}R
\label{15}
\end{equation}
is the Einstein tensor,
\begin{equation}
\nabla^{\rho}\nabla_{\rho} \varphi = 
\frac{1}{\sqrt{-g}}\partial^{\rho}\left( \sqrt{-g} \partial_{\rho}
\varphi \right)
\label{16}
\end{equation} 
is the generally covariant d'Alembertian for a scalar field, and
\begin{equation}
T_{\mu\nu} = \frac{2}{\sqrt{-g}} \frac{\delta \mathcal{L}_{m}}
{\delta g^{\mu\nu}}
\label{17}
\end{equation}
is the matter energy-momentum tensor.

Taking the trace of (\ref{12}), we find
\begin{equation}
6\varphi\left(\nabla^{\mu}\nabla_{\mu} - \frac{1}{6}R \right) \varphi 
= \frac{1}{2} k^2T,
\label{18}
\end{equation}
where $T = g^{\mu\nu}T_{\mu\nu}$. The field equations (\ref{13}) and 
(\ref{18}) require that $T=0$, which means that MCG couples consistently only 
to conformally invariant matter fields. This is not a problem because it is 
well known that all matter fields are conformally invariant in the Standard 
Model of particle physics. The massless matter fields are naturally conformally 
invariant whereas the massive matter fields become conformally invariant 
after the introduction of the Higgs field.


\section{The particle content}
\label{sec3}


Using the weak-field approximations
\begin{equation}
g_{\mu\nu} = \eta_{\mu\nu} + kh_{\mu\nu}, 
\label{19}
\end{equation}
\begin{equation}
\varphi = 1 + k\sigma,
\label{20}
\end{equation}
and keeping only the terms of second order in the fields $h_{\mu\nu}$ and 
$\sigma$, we find that (\ref{11}) reduces to the total linearized MCG action
\begin{eqnarray}
\bar{S}_{\textrm{tot}} &=&  \int{d^{4}x} \bigg[ 
\bar{\mathcal{L}}_{EH}(h_{\mu\nu}) + 2\sigma\bar{R} + 6 \partial^{\mu}\sigma
\partial_{\mu}\sigma   - \frac{1}{m^2}\left(\bar{R}^{\mu\nu}\bar{R}_{\mu\nu} 
- \frac{1}{3}\bar{R}^{2} \right)  \bigg] \nonumber \\ &&  
+ \int{d^{4}x}\left[\frac{1}{2}kh^{\mu\nu}\mathring{T}^{T}_{\mu\nu}\right],   
\label{21}
\end{eqnarray}
where $\mathring{T}^{T}_{\mu\nu}$ is the traceless part of the flat 
fermion energy-momentum tensor (see Appendix \ref{sec7}),
\begin{equation}
\bar{R}_{\mu\nu} = \frac{1}{2} \left( \partial_{\mu}\partial^{\rho}
h_{\rho\nu} + \partial_{\nu}\partial^{\rho}h_{\rho\mu} 
- \partial^{\rho}\partial_{\rho}h_{\mu\nu} 
- \partial_{\mu}\partial_{\nu}h  \right)
\label{22}
\end{equation}
is the linearized Ricci tensor,
\begin{equation}
\bar{R} =  \partial^{\mu}\partial^{\nu}h_{\mu\nu} 
- \partial^{\mu}\partial_{\mu}h
\label{23}
\end{equation} 
is the linearized scalar curvature, and
\begin{equation}
\bar{\mathcal{L}}_{EH}(h_{\mu\nu}) = - \frac{1}{4} \Big( \partial^{\rho}
h^{\mu\nu}\partial_{\rho}h_{\mu\nu} - 2\partial^{\mu}h^{\nu\rho}
\partial_{\rho}h_{\mu\nu}
+ 2\partial^{\mu}h_{\mu\nu}\partial^{\nu}h 
- \partial^{\mu}h\partial_{\mu}h  \Big)
\label{24}
\end{equation}
is the linearized Einstein-Hilbert Lagrangian density, with 
$h = \eta^{\mu\nu}h_{\mu\nu}$. 

In order to obtain a second-order derivative form, we choose the method 
of the decomposition into oscillator variables \cite{Pais} and write 
(\ref{21}) as
\begin{eqnarray}
\bar{S}_{\textrm{tot}} &=& \int{d^{4}x} \bigg[\frac{1}{4}h^{\mu\nu}
\Box q_{\mu\nu} - \frac{1}{2}h^{\mu\nu}\partial_{\mu}\partial^{\rho}
q_{\rho\nu} + \frac{1}{4}h^{\mu\nu}\partial_{\mu}\partial_{\nu}q
+\frac{1}{4}h\partial^{\mu}\partial^{\nu}q_{\mu\nu} \nonumber \\ && 
- \frac{1}{4}h \Box q  + \frac{m^2}{16}h^{\mu\nu}h_{\mu\nu} 
- \frac{m^2}{8}h^{\mu\nu}q_{\mu\nu} + \frac{m^2}{16}q^{\mu\nu}
q_{\mu\nu} - \frac{m^2}{16}h^{2}  + \frac{m^2}{8}hq \nonumber \\ &&  
- \frac{m^2}{16}q^{2} + 2\left( h^{\mu\nu} \partial_{\mu}
\partial_{\nu}\sigma - h \Box \sigma\right) - 6 \sigma\Box\sigma 
+ \frac{1}{2}kh^{\mu\nu}\mathring{T}^{T}_{\mu\nu} \bigg],
\label{25}
\end{eqnarray}
where  $\Box = \eta^{\mu\nu}\partial_{\mu}\partial_{\nu}$ and 
$q = \eta^{\mu\nu}q_{\mu\nu}$. Varying this action
with respect to $q^{\mu\nu}$ gives\footnote{The parenthesis in the 
indices denote symmetrization.}
\begin{equation}
q_{\mu\nu} = h_{\mu\nu} - \frac{2}{m^{2}} \Big[\Box h_{\mu\nu} 
- 2\partial^{\rho}\partial_{(\mu}h_{\nu)\rho} 
+ \partial_{\mu}\partial_{\nu}h  
+\frac{1}{3}\eta_{\mu\nu}\partial^{\rho}\partial^{\sigma}h_{\rho\sigma} 
-\frac{1}{3}\eta_{\mu\nu}\Box h  \Big] ,
\label{26}
\end{equation}
and with this the field equations obtained from (\ref{25}) are 
equivalent to the field equations obtained from (\ref{21}). Finally, 
with the change of variables
\begin{equation}
h_{\mu\nu} = A_{\mu\nu} + B_{\mu\nu},
\label{27}
\end{equation}
\begin{equation}
q_{\mu\nu} = A_{\mu\nu} - B_{\mu\nu},
\label{28}
\end{equation}
we find the action
\begin{eqnarray}
\bar{S}_{\textrm{tot}} &=& \int{d^{4}x} \bigg[\bar{\mathcal{L}}_{EH}
(A_{\mu\nu}) - \bar{\mathcal{L}}_{EH}(B_{\mu\nu}) + \frac{m^2}{4}\left( 
B^{\mu\nu}B_{\mu\nu} - B^{2} \right) \nonumber \\ && + 2\left( A^{\mu\nu}
\partial_{\mu}\partial_{\nu}\sigma- A\Box\sigma + B^{\mu\nu}\partial_{\mu}
\partial_{\nu}\sigma - B\Box\sigma\right) - 6 \sigma\Box\sigma 
\nonumber \\ &&  + \frac{1}{2}kA^{\mu\nu}\mathring{T}^{T}_{\mu\nu} 
+ \frac{1}{2}kB^{\mu\nu}\mathring{T}^{T}_{\mu\nu} \bigg],
\label{29}
\end{eqnarray}
where $A = \eta^{\mu\nu}A_{\mu\nu}$ and $B = \eta^{\mu\nu}B_{\mu\nu}$. 

In order to reveal the complete particle content of (\ref{29}), we 
consider the transformations
\begin{equation}
A_{\mu\nu} = A'_ {\mu\nu} + \frac{1}{m}\Big(\partial_{\mu}V_{\nu} 
+ \partial_{\nu}V_{\mu} \Big) + 2\eta_{\mu\nu}\sigma,
\label{30}
\end{equation} 
\begin{equation}
B_{\mu\nu} = B'_{\mu\nu} - \frac{1}{m}\Big(\partial_{\mu}V_{\nu} 
+ \partial_{\nu}V_{\mu} \Big) - 2\eta_{\mu\nu}\sigma,
\label{31}
\end{equation}
which gives
\begin{eqnarray}
\bar{S}_{\textrm{tot}} &=& \int{d^{4}x} \bigg[\bar{\mathcal{L}}_{EH}
(A'_{\mu\nu}) - \bar{\mathcal{L}}_{EH}(B'_{\mu\nu})  + \frac{m^2}{4}\left( 
B'^{\mu\nu}B'_{\mu\nu} - B'^{2} \right) + \frac{1}{4}F^{\mu\nu}F_{\mu\nu} 
\nonumber \\ && +3\left( m^{2} B'\sigma -2m\sigma\partial_{\mu}V^{\mu}\right)   
- m\left( B'^{\mu\nu}\partial_{\mu}V_{\nu} - B'\partial_{\mu}V^{\mu} \right)	
\nonumber \\ && - 6 \sigma\left(\Box +2m^2\right)\sigma 
+ \frac{1}{2}kA'^{\mu\nu}\mathring{T}^{T}_{\mu\nu} 
+ \frac{1}{2}kB'^{\mu\nu}\mathring{T}^{T}_{\mu\nu} \bigg],
\label{32}
\end{eqnarray}
where $F_{\mu\nu} = \partial_{\mu}V_{\nu} - \partial_{\nu}V_{\mu}$. 

Taking into account the conservation and the traceless condition of 
$\mathring{T}^{T}_{\mu\nu}$, we can see that (\ref{32}) is invariant 
under the gauge transformations 
\begin{equation}
A'_{\mu\nu} \rightarrow A'_{\mu\nu} + \partial_{\mu}\xi_{\nu} 
+ \partial_{\nu}\xi_{\mu},
\label{33}
\end{equation}
\begin{equation}
B'_{\mu\nu} \rightarrow B'_{\mu\nu} + \partial_{\mu}\chi_{\nu} + 
\partial_{\nu}\chi_{\mu}, \ \ \ V_{\mu} \rightarrow V_{\mu} 
+ m\chi_{\mu},
\label{34}
\end{equation}
\begin{equation}
B'_{\mu\nu} \rightarrow B'_{\mu\nu} + 2m\zeta\eta_{\mu\nu}, \ \  
V_{\mu} \rightarrow V_{\mu} + \partial_{\mu}\zeta, \ \  \sigma \rightarrow 
\sigma - m\zeta,
\label{35}
\end{equation}
where $\xi_{\mu}$ and $\chi_{\mu}$ are arbitrary spacetime dependent vector 
fields, and $\zeta$ is an arbitrary spacetime dependent scalar field. 
Classically, we may impose the gauge conditions
\begin{equation}
\partial^{\mu}A'_{\mu\nu} - \frac{1}{2}\partial_{\nu}A' = 0,
\label{36}
\end{equation}
\begin{equation}
\partial^{\mu}B'_{\mu\nu} - \frac{1}{2}\partial_{\nu}B' - mV_{\nu} = 0,
\label{37}
\end{equation}
\begin{equation}
\partial^{\mu}V_{\mu} - m\left(\frac{1}{2}B' + 3\sigma \right) = 0,
\label{38}
\end{equation}
which fix the gauge freedoms up to residual gauge parameters satisfying the 
subsidiary conditions
\begin{equation}
\Box \xi_{\mu} = 0,
\label{39}
\end{equation}
\begin{equation}
(\Box-m^2)\chi_{\mu} = 0,
\label{40}
\end{equation}
\begin{equation}
(\Box-m^2)\zeta = 0.
\label{41}
\end{equation}

By substituting (\ref{36})-(\ref{38}) into (\ref{32}), and integrating by 
parts, we obtain the diagonalized action
\begin{eqnarray}
\bar{S}^{d}_{\textrm{tot}} &=& \int{d^{4}x} 
\bigg[ \frac{1}{4}A'^{\mu\nu}\Box A'_{\mu\nu} - \frac{1}{8} A'\Box A' 
-\frac{1}{4}B'^{\mu\nu}\left(\Box - m^{2}\right)B'_{\mu\nu}  \nonumber \\ &&   
+ \frac{1}{8} B'\left(\Box - m^{2}\right)B' - \frac{1}{2}V^{\mu}\left(\Box 
- m^{2}\right)V_{\mu} - 6\sigma\left(\Box - m^{2}\right)\sigma \nonumber \\ &&
+ \frac{1}{2}kA'^{\mu\nu}\mathring{T}^{T}_{\mu\nu} 
+ \frac{1}{2}kB'^{\mu\nu}\mathring{T}^{T}_{\mu\nu} \bigg],
\label{42}
\end{eqnarray}
which is dynamically equivalent to action (\ref{21}). This action contains a 
positive energy massless tensor field $A'_{\mu\nu}$, a negative energy massive 
tensor field $B'_{\mu\nu}$, a negative energy massive vector field $V_{\mu}$, 
and a negative energy massive scalar field $\sigma$. In order to identify
which of these fields have physical degrees of freedom, we note that 
$A'_{\mu\nu}$ has $10$ degrees of freedom, $B'_{\mu\nu}$ has $10$ 
degrees of freedom, $V_{\mu}$ has $4$ degrees of freedom, and $\sigma$ has 
$1$ degree of freedom, giving a total of 25 ($=10+10+4+1$) degrees of freedom. 
The conditions (\ref{36})-(\ref{41}) reduce the degrees of freedom by 
$18$ ($=4+4+1+4+4+1$). Therefore, MCG has only $7$ ($=25-18$) physical 
degrees of freedom, $2$ for the massless field $A'_{\mu\nu}$ and $5$ for the 
massive field $B'_{\mu\nu}$.


\section{The massless limit}
\label{sec4}


In the massless limit $m \rightarrow 0$, the action (\ref{32}) reduces to
\begin{eqnarray}
\bar{S}_{\textrm{tot}} &=& \int{d^{4}x} \bigg[\bar{\mathcal{L}}_{EH}
(A'_{\mu\nu}) - \bar{\mathcal{L}}_{EH}(B'_{\mu\nu})  - 6 \sigma\Box\sigma 
 + \frac{1}{4}F^{\mu\nu}F_{\mu\nu} \nonumber \\ &&
+ \frac{1}{2}kA'^{\mu\nu}\mathring{T}^{T}_{\mu\nu} 
+ \frac{1}{2}kB'^{\mu\nu}\mathring{T}^{T}_{\mu\nu} \bigg].
\label{43}
\end{eqnarray}
It is not difficult to see that this action is invariant under the gauge 
transformations 
\begin{equation}
A'_{\mu\nu} \rightarrow A'_{\mu\nu} + \partial_{\mu}\xi_{\nu} 
+ \partial_{\nu}\xi_{\mu},
\label{44}
\end{equation}
\begin{equation}
B'_{\mu\nu} \rightarrow B'_{\mu\nu} + \partial_{\mu}\chi_{\nu} 
+ \partial_{\nu}\chi_{\mu},
\label{45}
\end{equation} 
\begin{equation}
V_{\mu} \rightarrow V_{\mu} 
+ \partial_{\mu}\zeta.
\label{46}
\end{equation}
Thus, by imposing the gauge conditions
\begin{equation}
\partial^{\mu}A'_{\mu\nu} - \frac{1}{2}\partial_{\nu}A' = 0,
\label{47}
\end{equation}
\begin{equation}
\partial^{\mu}B'_{\mu\nu} - \frac{1}{2}\partial_{\nu}B' = 0, 
\label{48}
\end{equation}
\begin{equation}
\partial^{\mu}V_{\mu} = 0,
\label{49}
\end{equation}
to (\ref{43}), and integrating by parts, we obtain the diagonalized action
\begin{eqnarray}
\bar{S}^{d}_{\textrm{tot}} &=& \int{d^{4}x} 
\bigg[ \frac{1}{4}A'^{\mu\nu}\Box A'_{\mu\nu} - \frac{1}{8} A'\Box A' 
-\frac{1}{4}B'^{\mu\nu}\Box B'_{\mu\nu} + \frac{1}{8} B'\Box B' 
\nonumber \\ && - \frac{1}{2}V^{\mu}\Box V_{\mu}   - 6\sigma\Box\sigma
+ \frac{1}{2}kA'^{\mu\nu}\mathring{T}^{T}_{\mu\nu} 
+ \frac{1}{2}kB'^{\mu\nu}\mathring{T}^{T}_{\mu\nu} \bigg],
\label{50}
\end{eqnarray}
which contains a positive energy massless tensor field $A'_{\mu\nu}$ with 
$10$ degrees of freedom, a negative energy massless tensor field $B'_{\mu\nu}$ 
with $10$ degrees of freedom, a negative energy massless vector field $V_{\mu}$ 
with $4$ degrees of freedom, and a negative energy massless scalar field 
$\sigma$ with $1$ degree of freedom.

The gauge conditions (\ref{47})-(\ref{49}) fix the gauge freedoms up to residual 
gauge parameters satisfying the subsidiary conditions
\begin{equation}
\Box \xi_{\mu} = 0,
\label{51}
\end{equation}
\begin{equation}
\Box\chi_{\mu} = 0,
\label{52}
\end{equation}
\begin{equation}
\Box \zeta = 0.
\label{53}
\end{equation}
The conditions (\ref{47})-(\ref{49}) and (\ref{51})-(\ref{53}) reduce the 
$25$ degrees of freedom of the MCG massless limit by $18$, giving a total of 
$7$ physical degrees of freedom, $2$ for the massles tensor field 
$A'_{\mu\nu}$, $2$ for the massless tensor field $B'_{\mu\nu}$, $2$ for the 
massless vector field $V_{\mu}$, and $1$ for the massless scalar field 
$\sigma$. Hence, the number of physical degrees of freedom is preserved in the 
massless limit of the theory. Two of the five degrees of freedom of the massive 
tensor field go into the massless vector field, and one goes into the massless 
scalar field. Since the scalar and vector fields don't couple to the source, 
MCG is free of the van Dam-Veltman-Zakharov (vDVZ) discontinuity \cite{vDV,Z}.

It is worth noticing that MCG differs from theories of gravity with only 
a massive tensor field, for which general relativity (GR) should be 
recovered in the massless limit. Usually is the absence of this 
recovery that is associated with the vDVZ discontinuity. However, the real 
meaning of the vDVZ discontinuity is that the massless limit of a massive 
theory is not equivalent to the massless theory, which violates the 
physical continuity principle \cite{Boulware}. That is why is so important to 
a massive theory to be free from the vDVZ discontinuity.

Recently, it was proposed that the vDVZ discontinuity of a local gravitational 
theory is closely related to the finiteness of the gravitational potential at 
the origin \cite{Myung2}. If this proposal is valid then the MCG potential 
should have a singularity at the origin, which we will show that is not true 
in the next section.


\section{The Newtonian singularity}
\label{sec5}


The variation of (\ref{42}) with respect to $A'^{\mu\nu}$ and $B'^{\mu\nu}$ 
gives the field equations
\begin{equation}
\Box \left(A'_{\mu\nu} -\frac{1}{2}\eta_{\mu\nu}A'\right)
= - k\mathring{T}^{T}_{\mu\nu},
\label{54}
\end{equation}
\begin{equation}
\left(\Box - m^{2}\right)\left(B'_{\mu\nu} -\frac{1}{2}\eta_{\mu\nu}B'\right)
= k\mathring{T}^{T}_{\mu\nu}.
\label{55}
\end{equation}
Taking the traces of (\ref{54}) and (\ref{55}), and replacing them back, we 
obtain
\begin{equation}
\Box A'_{\mu\nu}
= - k\mathring{T}^{T}_{\mu\nu},
\label{56}
\end{equation} 
\begin{equation}
\left(\Box - m^{2}\right)B'_{\mu\nu}
= k\mathring{T}^{T}_{\mu\nu},
\label{57}
\end{equation}
whose general solutions are given by
\begin{equation}
A'_{\mu\nu} = k \int \frac{d^{4}p}{(2\pi)^4}\frac{e^{ip\cdot x}}{p^{2}} \, 
\mathring{T}^{T}_{\mu\nu}(p),
\label{58}
\end{equation}
\begin{equation}
B'_{\mu\nu} = - k \int \frac{d^{4}p}{(2\pi)^4}\frac{e^{ip\cdot x}}{p^{2}+m^2} 
\, \mathring{T}^{T}_{\mu\nu}(p),
\label{59}
\end{equation}
where
\begin{equation}
\mathring{T}^{T}_{\mu\nu}(p) = \int d^{4}x\,e^{-ip\cdot x}\mathring{T}^{T}_{\mu\nu}(x)
\label{60}
\end{equation}
is the Fourier transform of the traceless part of the source.

In the case of a point particle source with mass $M$ at rest at the origin, 
for which $\mathring{T}^{f}_{\mu\nu}(x) = M \delta^{0}_{\mu}\delta^{0}_{\nu}
\delta^3(\textbf{x})$, we have
\begin{equation}
\mathring{T}^{T}_{\mu\nu} = \left(M \delta^{0}_{\mu}\delta^{0}_{\nu} 
+  \frac{1}{4}\eta_{\mu\nu}M\right) \delta^3(\textbf{x}),
\label{61}
\end{equation}
where we consider $\eta_{\mu\nu} = \mathrm{diag}( -1, +1, +1, +1)$.
Substituting the $00$ component of (\ref{61}) into (\ref{58}) and 
(\ref{59}), we find the general solutions
\begin{equation}
A'_{00} = k\frac{3M}{4}\int \frac{d^{3}\textbf{p}}{(2\pi)^4}
\frac{e^{i\textbf{p}\cdot \textbf{x}}}{\textbf{p}^{2}} = k\frac{3M}{4}
\frac{1}{4\pi r},
\label{62}
\end{equation}
\begin{equation}
B'_{00} = -k\frac{3M}{4} \int \frac{d^{3}\textbf{p}}{(2\pi)^4}
\frac{e^{i\textbf{p}\cdot \textbf{x}}}{\textbf{p}^{2}+m^2} 
= - k\frac{3M}{4}\frac{e^{-mr}}{4\pi r}.
\label{63}
\end{equation}
It then follows from (\ref{27}), (\ref{30}) and (\ref{31}) that
\begin{equation}
h_{00} = A'_{00} + B'_{00} = k\frac{3M}{16\pi r}\left( 1 - e^{-mr} \right).
\label{64}
\end{equation}
Finally, noting that $2\phi = -k h_{00}$, we find the MCG potential
\begin{equation}
\phi(r) = - \frac{GM}{r}\left( 1 - e^{-mr} \right),
\label{65}
\end{equation}
where we chose $k^{2} = 32\pi G/3$ in order to (\ref{65}) agree with the 
Newtonian potential in the limit where $m$ tend to infinity. It is not 
difficult to see that (\ref{65}) is finite at the origin, which is a necessary 
condition for the renormalizability of the theory \cite{Gia}. In addition, 
the massless limit provides a zero order term of $e^{-mr} \approx 1$ in the 
MCG potential, which confirms that the theory is free of the vDVZ discontinuity.


\section{Final remarks}
\label{sec6}


We have found that MCG contains only two linearized fields with physical 
degrees of freedom: the usual positive energy massless tensor field and a 
negative energy massive tensor field. In the massless limit, the number of 
physical degrees of freedom of the theory remains the same, with the massive 
tensor field being decomposed into massless tensor, vector and scalar fields. 
The absence of coupling between the vector and scalar fields and the source 
makes the theory free of the vDVZ discontinuity. In addition, it was shown 
that the MCG potential is singularity free, which indicates that the 
vDVZ discontinuity is not directly related to the finiteness of the 
gravitational potential at the origin.


\appendix
\section{Linearized MCG matter Lagrangian density}
\label{sec7}


As we saw in Sec. \ref{sec2}, the matter Lagrangian density has to be 
conformally invariant in MCG. For simplicity, let us consider the Lagangian 
density for a fermion field $\psi$ conformally coupled to the gravitational 
field $g_{\mu\nu}$, which is given by 
\begin{equation}
\mathcal{L}_{m} = -\sqrt{-g}\Bigg[
\frac{1}{12}S^{2}R + \frac{1}{2}\partial^{\mu}S\partial_{\mu}S 
+ \frac{1}{4!}\lambda S^{4} 
+ \frac{i}{2}\left(\, \overline{\psi}\gamma^{\mu}
D_{\mu}\psi - D_{\mu}\overline{\psi}\gamma^{\mu}\psi \right) + 
\mu S\overline{\psi}\psi\Bigg],
\label{66}
\end{equation}
where $S$ is a real Higgs scalar field, $\lambda$ and $\mu$ are dimensionless 
coupling constants, $\overline{\psi} = \psi^{\dagger}\gamma^{0}$ is the adjoint 
fermion field, $D_{\mu} = \partial_{\mu} + [\gamma^{\nu},\partial_{\mu}
\gamma_{\nu}]/8 - [\gamma^{\nu},\gamma_{\lambda}]
\Gamma^{\lambda}\,\!\!_{\mu\nu}/8$, and $\gamma^{\mu}$ are the general 
relativistic Dirac matrices, which satisfy the anticommutation relation 
$\{\gamma^{\mu},\gamma^{\nu}\} = 2g^{\mu\nu}$. 

The variation of (\ref{66}) with respect to the matter fields $S$, 
$\overline{\psi}$ and $\psi$ gives
\begin{equation}
\left(\nabla^{\mu}\nabla_{\mu} - \frac{1}{6}R \right) S 
+ \frac{1}{6}\lambda S^3 + \mu \overline{\psi}\psi = 0,
\label{67}
\end{equation}
\begin{equation}
i\gamma^{\mu}D_{\mu}\psi + \mu S \psi = 0,
\label{68}
\end{equation}
\begin{equation}
iD_{\mu}\overline{\psi}\gamma^{\mu} - \mu S \overline{\psi} = 0.
\label{69}
\end{equation}
By substituting (\ref{66}) into (\ref{17}), and using the matter field 
equations (\ref{67})-(\ref{69}), we can write the MCG matter energy-momentum 
tensor in the form
\begin{eqnarray}
T_{\mu\nu} &=& T^{f}_{\mu\nu} - \frac{1}{4}g_{\mu\nu}T^{f}  
+ \frac{1}{6}S^{2}\left(R_{\mu\nu} - \frac{1}{4}g_{\mu\nu}R\right) 
+ \frac{2}{3}\nabla_{\mu}S\nabla_{\nu}S \nonumber \\ && 
- \frac{1}{6}g_{\mu\nu}\nabla^{\rho}S\nabla_{\rho}S  
- \frac{1}{3} S\nabla_{\mu}\nabla_{\nu} S 
+ \frac{1}{12}g_{\mu\nu}S\nabla^{\rho}\nabla_{\rho} S,
\label{70}
\end{eqnarray}
where
\begin{equation}
T^{f}_{\mu\nu} = \frac{i}{4}\big(\, \overline{\psi}
\gamma_{\mu}D_{\nu}\psi - D_{\nu}\overline{\psi}\gamma_{\mu}\psi 
+ \overline{\psi}\gamma_{\nu}D_{\mu}\psi - D_{\mu}\overline{\psi}\gamma_{\nu}
\psi \big)
\label{71}
\end{equation}
is the fermion energy-momentum tensor, and 
\begin{equation}
T^{f} =  \frac{i}{2}\big(\, \overline{\psi}
\gamma^{\mu}D_{\mu}\psi - D_{\mu}\overline{\psi}\gamma^{\mu}\psi  \big)
\label{72}
\end{equation}
is the trace of the fermion energy-momentum tensor. 

With the help of the conformal transformations 
\begin{equation*}
\tilde{g}_{\mu\nu}=e^{2\theta(x)}\,g_{\mu\nu},
\ \ \ \ \
\tilde{S}=e^{-\theta(x)} S,
\end{equation*}
\begin{equation}
\tilde{\psi} = e^{-3\theta(x)/2}\,\psi,
\ \ \ \ \
\tilde{\gamma}^{\mu} = e^{-\theta(x)}\gamma^{\mu},
\label{73}
\end{equation}
its possible to verify that (\ref{66}) is conformally invariant. Using 
this symmetry, we can impose the unitary gauge $S = S_{0}$, where 
$S_{0}$ is a spontaneously broken constant expectation value for the Higgs 
field. In this case, (\ref{70}) reduces to
\begin{equation}
T_{\mu\nu} =  T_{\mu\nu}^{f} -  \frac{1}{4} g_{\mu\nu}T^f 
+ \frac{1}{6}S_{0}^{2}\left(R_{\mu\nu} - \frac{1}{4}g_{\mu\nu}R\right).
\label{74}
\end{equation}
It is not difficult to see that both (\ref{70}) and (\ref{74}) are traceless, 
which means that the traceless condition of the MCG matter energy-momentum is 
independent of the gauge. 

It is well known that the weak-field limit of (\ref{17}) leads to the 
linearized matter Lagrangian density
\begin{equation}
\bar{\mathcal{L}}_{m} = \frac{1}{2}kh^{\mu\nu}\mathring{T}_{\mu\nu},
\label{75}
\end{equation}
where $\mathring{T}_{\mu\nu}$ is the flat matter energy-momentum tensor. 
Taking the flat limit of (\ref{74}) and substituting into (\ref{75}), we find 
the linearized MCG matter Lagrangian density
\begin{equation}
\bar{\mathcal{L}}_{m} = \frac{1}{2}kh^{\mu\nu}\mathring{T}^{T}_{\mu\nu},
\label{76}
\end{equation}
 where
\begin{equation}
\mathring{T}^{T}_{\mu\nu} = \mathring{T}^{f}_{\mu\nu} - \frac{1}{4}\eta_{\mu\nu}
\mathring{T}^{f} 
\label{77}
\end{equation}
is the traceless part of the flat fermion energy-momentum tensor.


\begin{thebibliography}{99}

\bibitem{Faria1}
F. F. Faria, Eur. Phys. J. C \textbf{76}, 188 (2016).

\bibitem{Stelle1}
 K. S. Stelle, Gen. Relativ. Gravit. \textbf{9}, 353 (1978).

\bibitem{Smilga}
A. V. Smilga, Nucl. Phys. B \textbf{706}, 598 (2005).

\bibitem{Stelle2}
 K. S. Stelle, Phys. Rev. D \textbf{16}, 953 (1977).

\bibitem{Antoniadis}
I. Antoniadis, E.T. Tomboulis, Phys. Rev. D \textbf{33}, 2756 (1986).

\bibitem{Faria2}
F. F. Faria, Eur. Phys. J. C \textbf{77}, 11 (2017).

\bibitem{Faria3}
F. F. Faria, Eur. Phys. J. C \textbf{78}, 277 (2018).

\bibitem{Faria4}
F. F. Faria, Adv. High Energy Phys. \textbf{2014}, 520259 (2014).

\bibitem{Pais}
A. Pais and G. E. Uhlenbeck, Phys. Rev. \textbf{79}, 145 (1950). 

\bibitem{vDV}
H. van Dam and M. J. G. Veltman, Nucl. Phys. B \textbf{22}, 397 (1970).

\bibitem{Z}
V. I. Zakharov, JETP Letters (Sov. Phys.) \textbf{12}, 312 (1970).

\bibitem{Boulware}
D. G. Boulware and S. Deser, Phys. Rev. D \textbf{6}, 3368 (1972).

\bibitem{Myung2}
Y. S. Myung,  Phys. Rev. D \textbf{96}, 064026 (2017). 

\bibitem{Gia}
 B. L. Giacchini, Phys. Lett. B \textbf{766}, 306 (2017).


\end{thebibliography}
\end{document}